\documentclass[%
 reprint,
 aps,prl,
 twocolumn,superscriptaddress,notitlepage,nofootinbib,
 nolongbibliography
]{revtex4-2}

\usepackage{tabularx}
\usepackage{xcolor}
\usepackage{graphicx}
\usepackage{physics}
\usepackage{dcolumn}
\usepackage{bm}
\usepackage{multirow}

\usepackage[colorlinks,
            linkcolor=blue,
            anchorcolor=blue,
            citecolor=blue,
            urlcolor=blue]{hyperref}

\begin{document}

\title{Jet-Origin Identification and Its Application at an Electron-Positron Higgs Factory}

\author{Hao Liang}\thanks{These authors contribute equally to this work.}
\affiliation{
Institute of High Energy Physics, Chinese Academy of Sciences,\\
19B Yuquan Road, Shijingshan District, Beijing 100049, China
}
\affiliation{University of Chinese Academy of Sciences,\\
19A Yuquan Road, Shijingshan District, Beijing 100049, China}
\author{Yongfeng Zhu}\thanks{These authors contribute equally to this work.}
\affiliation{State Key Laboratory of Nuclear Physics and Technology, School of Physics, Peking University, Beijing, 100871, China}
\author{Yuexin Wang}
\affiliation{
Institute of High Energy Physics, Chinese Academy of Sciences,\\
19B Yuquan Road, Shijingshan District, Beijing 100049, China
}
\affiliation{China Center of Advanced Science and Technology, Beijing 100190, China}
\author{Yuzhi Che}
\affiliation{
Institute of High Energy Physics, Chinese Academy of Sciences,\\
19B Yuquan Road, Shijingshan District, Beijing 100049, China
}
\affiliation{University of Chinese Academy of Sciences,\\
19A Yuquan Road, Shijingshan District, Beijing 100049, China}
\author{Manqi Ruan}
\email{ruanmq@ihep.ac.cn}
\affiliation{
Institute of High Energy Physics, Chinese Academy of Sciences,\\
19B Yuquan Road, Shijingshan District, Beijing 100049, China
}
\affiliation{University of Chinese Academy of Sciences,\\
19A Yuquan Road, Shijingshan District, Beijing 100049, China}
\author{Chen Zhou}%
\email{czhouphy@pku.edu.cn}
\affiliation{State Key Laboratory of Nuclear Physics and Technology, School of Physics, Peking University, Beijing, 100871, China}
\author{Huilin Qu}
\email{huilin.qu@cern.ch}
\affiliation{
CERN, EP Department, CH-1211 Geneva 23, Switzerland
}

\date{\today}

\begin{abstract}

To enhance the scientific discovery power of high-energy collider experiments, 
we propose and realize the concept of jet origin identification that categorizes jets into 5 quark species $(b,c,s,u,d)$, 5 anti-quarks $(\bar{b},\bar{c},\bar{s},\bar{u},\bar{d})$, and the gluon. 
Using state-of-the-art algorithms and simulated $\nu\bar{\nu}H, H\rightarrow jj$ events at 240~GeV center-of-mass energy at the electron-positron Higgs factory,
the jet origin identification simultaneously reaches jet flavor tagging efficiencies ranging from 67\% to 92\% for bottom, charm, and strange quarks, and jet charge flip rates of 7\% to 24\% for all quark species.
We apply the jet origin identification to Higgs rare and exotic decay measurements at the nominal luminosity of the Circular Electron Positron Collider (CEPC), and conclude that the upper limits on the branching ratios of $H\rightarrow s \bar{s}, u\bar{u}, d\bar{d}$, and $H\rightarrow sb, db, uc, ds$ can be determined to
$2\!\!\times\!\!10^{-4}$ to $1\!\!\times\!\!10^{-3}$ at 95\% confidence level. 
The derived upper limit for $H\rightarrow s \bar{s}$ decay is approximately three times the prediction of the Standard Model.

\end{abstract}

\maketitle

\begin{figure}[!htbp]
    \centering
    \includegraphics[scale=0.3]{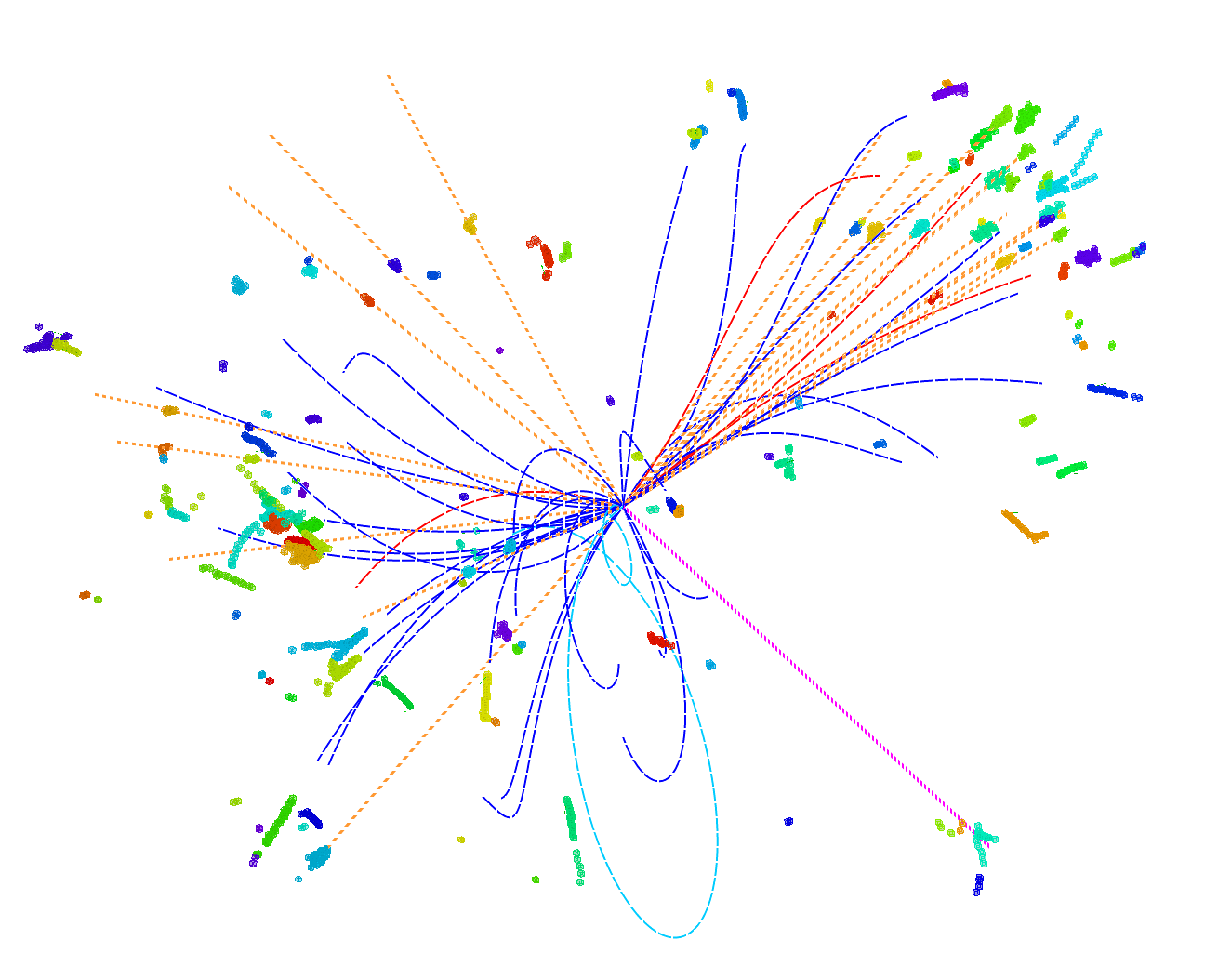}
    \caption{\label{fig:jets}Event display of an $e^+e^-\rightarrow \nu\bar{\nu} H\rightarrow \nu\bar{\nu} gg$ ($\sqrt{s}$ = 240 GeV) event simulated and reconstructed with the CEPC baseline detector~\cite{CEPC_CDR_Phy}.
    Different particles are depicted with colored curves and straight lines: \textcolor{red}{red} for $e^{\pm}$, \textcolor{cyan}{cyan} for $\mu^{\pm}$, \textcolor{blue}{blue} for $\pi^{\pm}$, \textcolor{orange}{orange} for photons, and \textcolor{magenta}{magenta} for neutral hadrons.}
\end{figure}

\emph{Introduction.---}
Quarks and gluons are standard model (SM) particles that carry color charges of the strong interaction.
Due to the color confinement of quantum chromodynamics (QCD), colored particles cannot travel freely in spacetime and are confined to composite particles like hadrons.
Once generated in high-energy collisions, quarks and gluons fragment into numerous particles that travel in directions approximately collinear to the initial colored particles.
These collinear particles are called jets, see Fig.~\ref{fig:jets}.

We define jet origin identification as the procedure to determine from which colored particle a jet is generated, and 
consider 11 different kinds: $b$, $\bar{b}$, $c$, $\bar{c}$, $s$, $\bar{s}$, $u$, $\bar{u}$, $d$, $\bar{d}$, and gluon.
A successful jet origin identification is critical for experimental particle physics at the energy frontier. 
At the Large Hadron Collider, successfully distinguishing quark jets from gluon ones could efficiently reduce the typically large background from QCD processes~\cite{Gallicchio:2011xq, LHCHiggsCrossSectionWorkingGroup:2016ypw, Gauld:2022lem, ATLAS:2018kot, CMS:2018nsn, CMS:2022psv, Metodiev:2018ftz}.
Jet flavor tagging is essential for the Higgs property measurements at the LHC~\cite{CMS:2018nsn, ATLAS:2020bhl, ATLAS:2022ers, CMS:2022psv}.
The determination of jet charge~\cite{Kang:2023ptt, Cui:2023kqb} was essential for weak mixing angle measurements at both LEP and LHC~\cite{CMS:2018ktx}, is critical for time-dependent CP measurements~\cite{Belle:2006dlp, LHCb:2019nin}, and could have a significant impact on Higgs boson property measurements~\cite{Li:2023tcr}.

We realize the concept of jet origin identification in physics events at an electron-positron Higgs factory using a Geant4-based simulation~\cite{GEANT4:2002zbu} (referred to as full simulation for simplicity), since the electron-positron Higgs factory is identified as the highest-priority future collider project~\cite{European:2720131, CEPCPhysicsStudyGroup:2022uwl}.
We develop the necessary software tools, Arbor~\cite{Arbor, ruan2018Reconstruction} and ParticleNet~\cite{Qu:2019gqs}, for the particle flow event reconstruction and the jet origin identification.
We demonstrate the jet origin identification performance using an 11-dimensional confusion matrix (referred to as $M_{11}$ for simplicity), which exhibits the performance of jet flavor tagging and jet charge measurements. 
We apply the jet origin identification to rare and exotic Higgs boson decay measurements under the CEPC nominal Higgs operation scenario. 
This scenario expects an integrated luminosity of 20~${\rm ab}^{-1}$ at $\sqrt{s}=240~{\rm GeV}$, and could accumulate 4 million Higgs bosons~\cite{CEPCPhysicsStudyGroup:2022uwl, Gao:2022lew}.
We analyze the rare decays $H\rightarrow  s\bar{s}$,  $u\bar{u}$, and $d\bar{d}$, and the flavor-changing neutral current (FCNC) decays $H\rightarrow sb$, $ds$, $db$, and $uc$ (here $sb$ denotes $s\bar{b}$ or $\bar{s}b$, and similarly for $ds$, $db$, and $uc$). 
We derive upper limits ranging from $10^{-3}$ to $10^{-4}$ for these seven processes. 
In the SM, the predicted branching ratio for the $H\rightarrow s\bar{s}$ process is $2.3\times 10^{-4}$~\cite{Duarte-Campderros:2018ouv} and the derived upper limit corresponds to three times the SM prediction. 
The branching ratios for $H\rightarrow u\bar{u}$ and $d\bar{d}$ are expected to be smaller than $10^{-6}$~\cite{ParticleDataGroup:2022pth, Duarte-Campderros:2018ouv, Denner:2011mq, Herren:2017osy}, while branching ratios for the above-mentioned FCNC processes are expected to be smaller than $10^{-7}$ from loop contributions~\cite{Kamenik:2023ytu}.

\emph{Detector Geometry and Software Tools.---}
We simulate $\nu\bar{\nu}H, H\rightarrow u\bar{u}$, $d\bar{d}$, $s\bar{s}$, $c\bar{c}$, $b\bar{b}$, and $gg$ processes at 240~GeV center-of-mass energy with the CEPC baseline detector~\cite{CEPC_CDR_Phy}.
The CEPC baseline detector design is a particle-flow-oriented concept composed of a high-precision vertex system, a large-volume gaseous tracker, high granularity calorimetry, and a large-volume solenoid.
We use Pythia-6.4~\cite{Pythia_6} for the event generations and MokkaPlus~\cite{MoradeFreitas:2004sq,Mokka_CEPCNote} for the Geant4-based detector simulation~\cite{GEANT4:2002zbu}.
The simulated samples are processed with the Arbor particle flow algorithm that
reconstructs all final-state particles and identifies their species.
The reconstructed final-state particles in a physics event are clustered into two jets using the $e^+e^-$-$k_t$ algorithm~\cite{Suehara:2015ura, Catani:1991hj}. 
For each jet, the kinematic and species information of all its final-state particles, including the track impact parameters associated with charged final-state particles, are input to a modified ParticleNet algorithm. 
The algorithm calculates the likelihoods corresponding to 11 different jet categories. 
For each process, one million physics events are simulated, where 600k events are used for training, 200k for validation, and 200k for testing. 
The model is trained for 30 epochs, and the epoch demonstrating the best accuracy on the validation sample is selected and applied to the testing sample to extract the numerical results. 



Information on the species of the final-state particles is critical for jet origin identification. We compare three scenarios to understand the impact of particle identification. 
The first scenario assumes perfect identification of charged leptons, i.e., $e^{\pm}$ and $\mu^{\pm}$ can be perfectly differentiated from each other and from charged hadrons.
The second scenario further assumes perfect identification of the species of charged hadrons (proton, anti-proton, $\pi^{\pm}$ and $K^{\pm}$). 
On top of the second scenario, the third one assumes perfect identification of $K_S^0$ and $K_L^0$.
For simplicity, the assignment of particle identification is based on MC truth.
On the other hand, full simulation performance studies show that the CEPC baseline detector could identify leptons with an efficiency of 99.5\% with a hadron-to-lepton misidentification rate below 1\%~\cite{Yu:2020bxh, Yu:2021pxc}. It could also distinguish different species of charged hadrons ($\pi^{\pm}$, $K^{\pm}$, proton, and anti-proton) to better than $2\sigma$~\cite{An:2018jtk, Zhu:2022hyy, Che:2022dig}
and reconstruct $K_S^0$ and $\Lambda$ with a typical efficiency (purity) of 80\% (90\%) if they decay into charged particles~\cite{KsLambda}. 
Therefore, the second scenario is used as the default one since it matches the CEPC baseline detector performance, while the third scenario is used for comparison as the $K_L^0$ identification remains challenging. 

\begin{figure}[!htbp]
    \centering
    \includegraphics[scale=0.55]{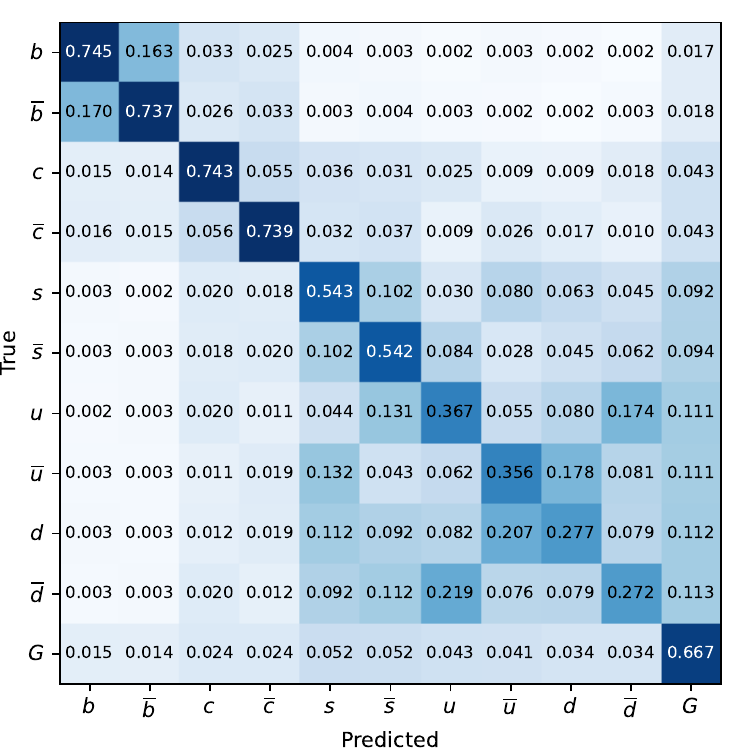}
    \caption{The confusion matrix $M_{11}$ with perfect identification of leptons and charged hadrons for $\nu \bar{\nu}H, H\rightarrow jj$ events at 240 GeV center-of-mass energy. The matrix is normalized to unity for each truth label (row).}
    \label{fig:M11}
\end{figure}

Figure~\ref{fig:M11} shows the overall jet origin identification performance with an 11-dimensional confusion matrix, $M_{11}$, derived by classifying each jet into the category with the highest likelihood.
In the quark sector, $M_{11}$ is approximately symmetric and block diagonalized into $2\times2$ blocks, corresponding to each specific species of quark. Meanwhile, gluon jets can be identified with an efficiency of 67\%.

\begin{figure}[!htbp]
    \centering
    \includegraphics[scale=0.55]{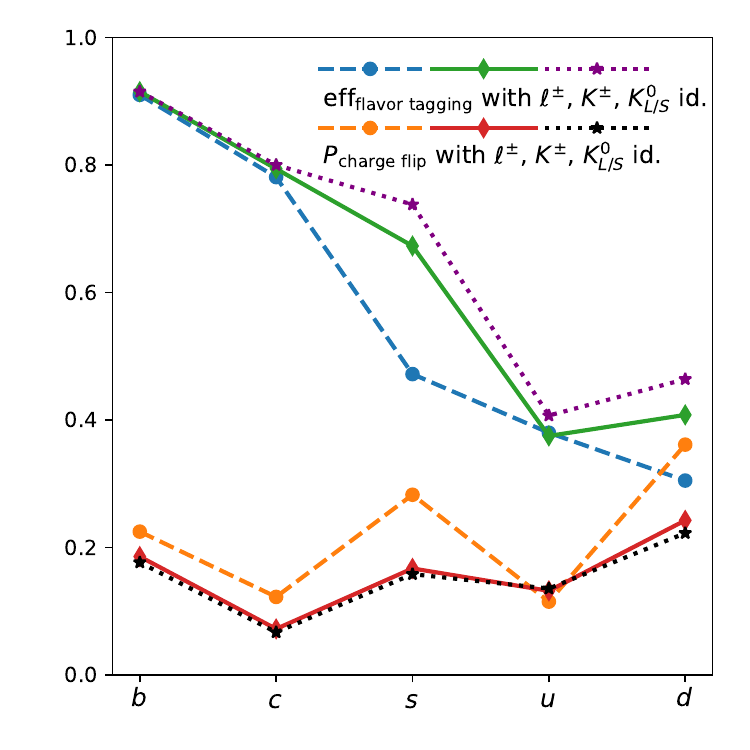}
    \caption{Jet flavor tagging efficiencies and charge flip rates with perfect identification of leptons (the first scenario, denoted as $\ell^{\pm}$ in the legend), plus identification of charged hadrons (the second and default scenario, denoted as $K^{\pm}$) and neutral kaons (the third scenario, denoted as $K^0_{L/S}$).}
    \label{fig:chargemisid}
\end{figure}

The performance of the jet origin identification can be studied in more detail via jet flavor tagging efficiencies and charge flip rates.
For each jet, we compare the gluon likelihood and the five sums of quark and anti-quark likelihoods of every kind.
The jet flavor is then defined as the kind with the highest value.
The jet charge is determined by comparing the likelihoods between the quark and the anti-quark. 
Figure~\ref{fig:chargemisid} illustrates the derived jet flavor tagging efficiencies and charge flip rates, which slightly differ from $M_{11}$~due to the different procedure described above.

Figure~\ref{fig:chargemisid} additionally compares the performance under different particle identification scenarios.
In the default scenario, represented by the solid lines, the $b$/$c$/$s$ jets could attain tagging efficiencies of 92\%/79\%/67\% and charge flip rates of 19\%/7\%/17\%, respectively.
The identification of $u$ and $d$ jets is less accurate, amounting to tagging efficiencies of 37\% to 41\% and jet charge flip rates of 13\% to 24\%. 
Noticeably, the down-type jets have a significantly higher jet charge flip rate than the up-type jets, since the latter carries twice the absolute charge as the former.
Of all types, the $c$ jets have the lowest charge flip rate as they are heavier and of the up-type.
Figure~\ref{fig:chargemisid} also exhibits the impact of final-state particle identification on jet origin identification. 
Compared to the scenario with only lepton identification, introducing charged hadron identification (the default scenario) enhances the $s$-tagging efficiency from 47\% to 67\%.
Concurrently, it reduces the jet charge flip rates across all types except for $u$.
Additionally, it significantly improves the $d$-tagging efficiency. 
The third scenario that includes neutral kaon information further enhances the $s$-tagging efficiency to 74\%.
However, the jet charge flip rates remain the same as in the second scenario, since $K_S^0$ and $K_L^0$ are superpositions of $\ket{s\bar{d}}$ and $\ket{\bar{s}d}$ states, meaning their identification has no impact on distinguishing quarks from anti-quarks.

\emph{Benchmark Physics Analyses.---}
The precise measurement of Higgs boson properties is a central objective for particle physics. 
The anticipated precision of Higgs measurements at future Higgs factories has been extensively studied, showing that the major SM decay modes can be measured with a relative accuracy of 0.1\% to 1\% at electron-positron Higgs factories~\cite{An:2018dwb, CEPCPhysicsStudyGroup:2022uwl, Asner:2013psa, Bernardi:2022hny},
surpassing the expected precision at the High Luminosity-LHC (HL-LHC) by one order of magnitude~\cite{Cepeda:2019klc}.
Meanwhile, the rare and FCNC decays of the Higgs boson are of great interest to many New Physics models~\cite{Duarte-Campderros:2018ouv,
Kamenik:2023ytu, Bejar:2004rz, Albert:2022mpk, CMS-PAS-SMP-22-006, Barducci:2017ioq}.

We explore the anticipated upper limits of $H\rightarrow s\bar{s}$, $u\bar{u}$,  $d\bar{d}$, and $H\rightarrow sb$, $ds$, $db$, $uc$ at the CEPC, where Higgs bosons are mainly produced via the Higgsstrahlung ($ZH$) and vector boson fusion  ($e^+e^-\rightarrow \nu_e\bar{\nu_e} H$, $e^+e^-\rightarrow e^+e^- H$) processes~\cite{Mo:2015mza}.
Our simulation analyses focus on the $\nu\bar{\nu} H$, $\mu^+\mu^- H$, and $e^+e^- H$ channels, with expected event yields of 0.926, 0.135, and 0.141 million under the CEPC nominal Higgs operation scenario, respectively.

We begin with the existing analyses of $\nu\bar{\nu} H, H\rightarrow b\bar{b}$,  $c\bar{c}$, $gg$~\cite{Zhu:2022lzv, Bai:2019qwd} at center-of-mass energy of 240 GeV.
These analyses consist of two stages: the first stage performs event selection to concentrate the Higgs to di-jet signal in the entire SM data sample, and the second stage identifies different flavor combinations using the LCFIPlus~\cite{Suehara:2015ura} flavor tagging algorithm.  
For the Higgs rare and exotic decay analyses, 
we re-optimize the event selections in the first stage and replace the flavor tagging in the second stage with the jet origin identification. 
After the event selections (described briefly in Appendix A), the leading SM backgrounds are mainly $\ell\bar{\nu}_{\ell} W$, $\nu\bar{\nu} Z$, and $\ell^+\ell^- Z$ events.
Taking the $\nu\bar{\nu} H, H\rightarrow jj$ analyses as an example, 
the event selection in this stage has a final signal efficiency of 24\%, and reduces the backgrounds by six orders of magnitude, leading to a background yield of 23k.
A toy MC simulation is then applied to the remaining events to mimic the jet origin identification, by sampling the $11$ likelihoods of each jet according to its origin. 
A gradient boosting decision tree (GBDT) classifier~\cite{ke2017lightgbm} is trained to distinguish signal and background processes using the $22$ likelihoods of the jet pair in a physics event.

\begin{figure}[!htbp]
    \centering
    \includegraphics[scale=0.62]{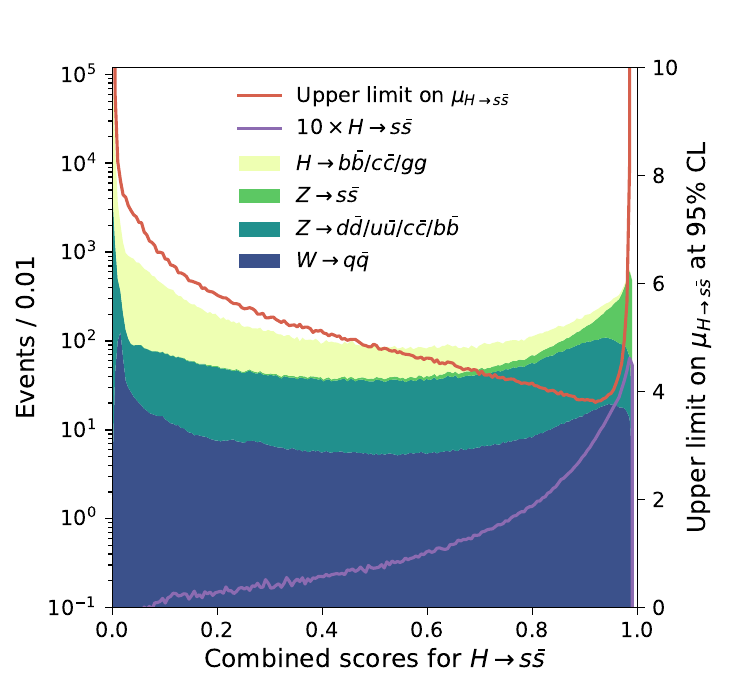} 
    \includegraphics[scale=0.62]{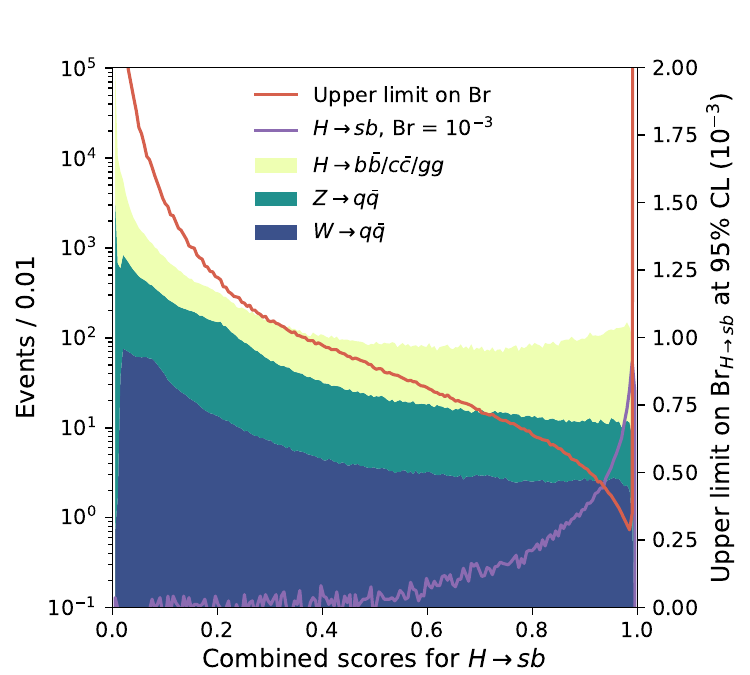}    
    \caption{The distributions of combined scores for signal and SM backgrounds,
    where the signals are (upper panel) $H\rightarrow s\bar{s}$ and (lower panel) $H\rightarrow sb$, respectively, in the $\nu\bar{\nu} H$ process, with CEPC nominal parameters.}
    \label{fig:HssDistribution}
\end{figure}

For the $\nu\bar{\nu} H, H\rightarrow s\bar{s}$ analysis, 
the combined GBDT scores of the remaining events are illustrated in the upper panel of Fig.~\ref{fig:HssDistribution}.
Defining the signal strength as the ratio of the observed event yield to the SM prediction,
the anticipated upper limit on the signal strength of $H\rightarrow s\bar{s}$ at 95\% confidence level (CL)~\cite{Read:2000ru, Read:2002hq} as a function of cut value is shown in Fig.~\ref{fig:HssDistribution}.
With the optimal cut on the combined scores, there remain 37 events of $H\rightarrow s\bar{s}$ and 5.1k background events, leading to an upper limit of 3.8 on the signal strength of $H\rightarrow s\bar{s}$ at 95\% CL.
A fit to the combined score distributions further improves the upper limit to 3.5.
Combined with $e^+e^-H$ and $\mu^+\mu^- H$ channels, an expected upper limit of 3.2 on the signal strength is achieved at 95\% CL.
It is worth noting that in the analysis of $H \rightarrow s\bar{s}$,  the branching ratios of all other Higgs decays are assumed to be at their SM predictions.

\begin{figure}[!htbp]
    \centering
    \includegraphics[scale=0.7]{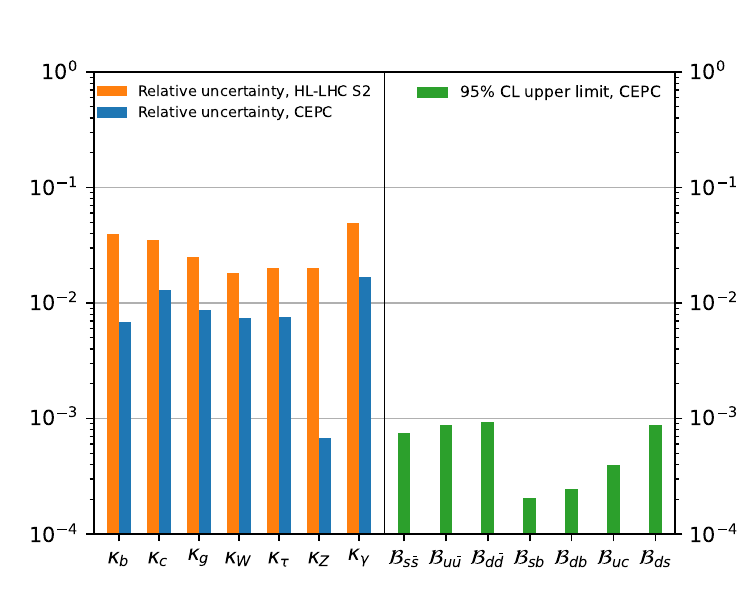}
    \caption{Expected upper limits on the branching ratios of rare Higgs boson decays from this work (green) and 
    the relative uncertainties of Higgs couplings anticipated at CEPC \cite{CEPCPhysicsStudyGroup:2022uwl} (blue) and HL-LHC~\cite{Cepeda:2019klc} (orange) under the kappa-0 fit scenario~\cite{deBlas:2019rxi} and scenario S2 of systematics~\cite{DeBlas:2019qco},
    as cited in Ref. \cite{CEPCPhysicsStudyGroup:2022uwl}. The limit on $B_{s\bar{s}}$ corresponds to an upper limit of 1.7 on the Higgs-strange coupling modifier $\kappa_s$ (not shown).
    }
    \label{fig:uplimit}
\end{figure}

We analyze $H\rightarrow u\bar{u}$ and $H\rightarrow d\bar{d}$ decay modes using the same method. 
By combining all three channels, the branching ratios of $H\to u\bar{u}$ and $d\bar{d}$
can be constrained to 0.091\% and 0.095\% at 95\% CL, respectively.
These results are less stringent than those for $H\to s\bar{s}$ since the identification of the $u$ and $d$ jets is much more challenging than $s$ jets.
We also analyze $H\rightarrow sb$, $ds$, $db$, and $uc$ decay modes and obtain upper limits ranging from 0.02\% to 0.1\% for these decay modes. 
These results are summarized in Table~\ref{tab:uplimits} and Fig.~\ref{fig:uplimit}.

\begin{table}[!htbp]
    \centering
    {\small
    \setlength{\tabcolsep}{2.5pt}
    \caption{
  Summary of background yields from $H\rightarrow b\bar{b}/c\bar{c}/gg$, $Z$, and $W$ prior to the flavor-based event selection, along with the expected upper limits on Higgs decay branching ratios at 95\% CL under the background-only hypothesis.
    }    
    \label{tab:uplimits}    
    
    \begin{tabular}{c|ccc|cccccccc}
    \hline
    &\multicolumn{3}{  c |}{ Bkg. ($10^3$) } & \multicolumn{7}{c}{ Upper limits on Br. ($10^{-3}$)} \\
             &$H$ & $Z$ & $W$ & $s\bar{s}$ & $u\bar{u}$ & $d\bar{d}$& $sb$   & $db$ &  $uc$  & $ds$\\
    \hline

    $\nu \bar{\nu} H$ &151 & 20 & 2.1 &0.81&0.95&0.99&0.26&0.27&0.46&0.93 \\
    $\mu^+ \mu^- H$ & 50& 25 &0  &2.6&3.0&3.2&0.5&0.6&1.0&3.0\\
    $e^+e^- H$ &26 & 16 & 0 &4.1&4.6&4.8&0.7&0.9&1.6&4.3 \\
    Comb. &- & - & - &0.75&0.91&0.95&0.22&0.23&0.39&0.86 \\
    \hline
    \end{tabular}
    
    }
\end{table}

\emph{Discussion and Summary.---}
We propose the concept of jet origin identification that distinguishes jets generated from 11 types of colored SM particles.
State-of-the-art algorithms are developed to realize the concept of jet origin identification at the future electron-positron Higgs factory, achieving jet flavor tagging efficiencies ranging from 67\% to 92\% for bottom, charm, and strange quarks, and jet charge flip rates of 7\% to 24\% for all species of quarks.

We analyze the impact of final-state particle identification on jet origin identification and find that charged hadron identification is critical for both jet flavor tagging and charge measurement. 
The identification of neutral kaons further enhances jet flavor tagging performance but has no impact on jet charge measurement, as expected.

Utilizing jet origin identification, we estimate the upper limits for seven rare and FCNC hadronic decay modes of the Higgs boson. 
We conclude that the branching ratios of these decay modes could be constrained to 0.02\% to 0.1\% at 95\% CL in the nominal CEPC Higgs operation scenario.
For the $H\rightarrow s\bar{s}$ decay, the expected upper limit is approximately three times the SM prediction, which improves by more than a factor of two upon previous studies~\cite{Duarte-Campderros:2018ouv, Albert:2022mpk}.
The improvement here is largely attributed to our state-of-the-art jet origin identification algorithm, which is capable of exploiting the information of all particles in a jet, not just the kaon particles.
The upper limits for $H\rightarrow u\bar{u}/d\bar{d}$ can be interpreted as constraints on the Higgs-quark couplings of $<101$ and $<37$ times the SM predictions, respectively (i.e. $\kappa_u < 101$ and $\kappa_d < 37$).
This improves upon existing analyses by roughly one order of magnitude.
Regarding the Higgs-boson FCNC decay, a previous study using DELPHES~\cite{deFavereau:2013fsa} fast simulation indicated that the branching fraction for $H\rightarrow sb$ ($H\rightarrow db$) could be constrained to $10^{-2}$ with an integrated luminosity of 30 ${\rm ab}^{-1}$~\cite{ILYUSHIN2020114921}, while our results show an improvement of two (one) order of magnitude.
We also quantify the upper limits for $H\rightarrow uc$ and $H\rightarrow ds$ in Table~\ref{tab:uplimits}.

Many systematic and theoretical uncertainties are relevant to jet origin identification, including detector performance, beam-induced backgrounds, the number of pile-up events, jet kinematics, jet clustering algorithms, hadronization models, etc.
Appendix B summarizes a series of relevant comparison studies. 
In short, we conclude that jet origin identification performance is stable with respect to jet kinematics in the relevant energy range, see Fig.~\ref{fig:energy}.
We observe that the performances obtained from hadronic $Z$ processes at 91.2~GeV and the $\nu\bar{\nu} H$ processes at 240 GeV are statistically consistent within the detector's fiducial region, as shown in Figs.~\ref{fig:ERHvsZ} and~\ref{fig:ERZH}.
In other words, the jet origin identification could be calibrated using the large number of events in the $Z$-pole sample to control the performance-relevant systematic uncertainties for the physics measurements including the Higgs property measurements.
We observe comparable performance for different hadronization models with small but visible differences, see Fig.~\ref{fig:P6H7}. 
These analyses lay the foundation for the application of jet origin identification at the energy frontier, especially in physics measurements with relatively larger statistical uncertainties, while more dedicated studies are certainly needed. 

The jet origin identification algorithm reads critical information from all the reconstructed particles and provides much higher separation power between jets stemming from different species of colored SM particles. 
Consequently, this could significantly enhance the scientific discovery potential for physics measurements with multi-jet final states, such as those expected at future Higgs factories.
Jet origin identification appreciates a detector capable of efficiently distinguishing final state particles and identifying their species information, as demonstrated in Fig.~\ref{fig:chargemisid}.
Recent studies also suggest that a light, precise vertex detector located close to the interaction point is favorable for jet origin identification \cite{Wu:2018awx,Zhu:2023xpk}.
Co-evolving with state-of-the-art detector technology, reconstruction algorithms, and Artificial Intelligence, the jet
origin identification algorithm developed here indicates that colored SM particles could potentially be identified with comparable performance to leptons and photons.

\begin{acknowledgments}
We would like to thank Christophe Grojean and Michele Selvaggi for the delightful discussions, and Qiang Li, Gang Li, Congqiao Li, Yuxuan Zhang, and Sitian Qian for their support with the software tools. We would also like to thank Xiaoyan Shen for her continuous supports.
This work is supported by the Innovative Scientific Program of the Institute of High Energy Physics, the National Natural Science Foundation of China under grant No.~12342502,
and the Fundamental Research Funds for the Central Universities, Peking University.
We appreciate the Computing Center at the Institute of High Energy Physics for providing the computing resources.
\end{acknowledgments}

\appendix
\section{\label{appA}Appendix A: Event selection of benchmark analyses}
This appendix describes the event selection for physics benchmark analyses presented in the letter.

We take as reference the existing full-simulation analysis of $\nu\bar{\nu} H, H \rightarrow b\bar{b}$, $c\bar{c}$, $gg$ at the CEPC~\cite{Zhu:2022lzv}.
This reference simulation analysis considers a nominal luminosity of 5.6 $\rm ab^{-1}$.
It includes all major SM backgrounds, with a total of $4.6\times10^{7}$ physics events simulated and processed using the CEPC baseline software, and concludes that the signal strength of the $\nu\bar{\nu} H, H \rightarrow b\bar{b}$, $c\bar{c}$, $gg$ processes can be measured with a relative precision of 0.49\%, 5.8\%, and 1.8\%, respectively.
 
All benchmark analyses of $\nu\bar{\nu} H, H \rightarrow jj$ in this letter use the same kinematic variables for the event selection as in the reference analysis. 
These kinematic variables include total recoil mass ($M_{\rm recoil}$), total visible mass ($M_{\rm invariant}$), total visible energy ($E_{\rm vis}$), total transverse momentum ($P_{T}$), energies of the leading lepton candidate and leading neutral particle, and the Durham distance $y_{23}$~\cite{ Catani:1991hj} that describes the event topology. 
A loose cut is applied to the sample, with an efficiency of 40\% on the $\nu\bar{\nu} H, H \rightarrow jj$ process and a reduction of the background to 495k. 
A BDT cut that combines these kinematic and topological variables is applied, which further suppresses the SM background to 23k and has an efficiency of 24\% on the $\nu\bar{\nu} H, H \rightarrow jj$ signal, see Table~\ref{tab:cutsFlow}.

\begin{table}[!htbp]
    \centering
    {\footnotesize
    \setlength{\tabcolsep}{2.5pt}
    \caption{The event selection of $\nu\bar{\nu} H(H\to q\bar{q}/gg)$ when CEPC operates as a Higgs factory at the center-of-mass energy of 240 GeV and collects an integrated luminosity of 20 ab$^{-1}$.
   The $\gamma\gamma$ label is the abbreviation of $\gamma\gamma\to$ hadrons process, and S$W$/S$Z$ refers to single W and single Z processes. The units for mass, energy, and momentum are ${\rm GeV}/c^2$, ${\rm GeV}$, and ${\rm GeV}/c$, respectively.}    
    \label{tab:cutsFlow}    
\setlength{\tabcolsep}{2pt}

    \begin{tabular}{ccccccc}
      \hline
                                 & $\nu\bar{\nu} H q\bar{q}/gg$               & 2$f$/$\gamma\gamma$                        & S$W$/S$Z$                                  & $WW$/$ZZ$                                  & $ZH$                                       \\
      \hline
      \multirow{2}{*}{Total}     & \multirow{2}{*}{$6.4\!\!\times\!\!10^{5}$} & \multirow{2}{*}{$4.6\!\!\times\!\!10^{9}$} & \multirow{2}{*}{$1.1\!\!\times\!\!10^{8}$} & \multirow{2}{*}{$2.8\!\!\times\!\!10^{8}$} & \multirow{2}{*}{$3.4\!\!\times\!\!10^6$}   \\
                                 &                                            &                                            &                                            &                                            &                                            \\
      $M_{\rm recoil}$           & \multirow{2}{*}{$5.6\!\!\times\!\!10^{5}$} & \multirow{2}{*}{$2.8\!\!\times\!\!10^{8}$} & \multirow{2}{*}{$1.4\!\!\times\!\!10^{7}$} & \multirow{2}{*}{$2.4\!\!\times\!\!10^{7}$} & \multirow{2}{*}{$2.7\!\!\times\!\!10^{5}$} \\
      $\in(74, 131)$             &                                            &                                            &                                            &                                            &                                            \\
      $E_{\rm vis}$              & \multirow{2}{*}{$5.1\!\!\times\!\!10^{5}$} & \multirow{2}{*}{$1.3\!\!\times\!\!10^{8}$} & \multirow{2}{*}{$8.8\!\!\times\!\!10^{6}$} & \multirow{2}{*}{$6.4\!\!\times\!\!10^{6}$} & \multirow{2}{*}{$1.8\!\!\times\!\!10^{5}$} \\
      $\in(109, 143)$            &                                            &                                            &                                            &                                            &                                            \\
      $E_{\rm leading\ lepton}$  & \multirow{2}{*}{$5.1\!\!\times\!\!10^{5}$} & \multirow{2}{*}{$1.2\!\!\times\!\!10^{8}$} & \multirow{2}{*}{$4.0\!\!\times\!\!10^{6}$} & \multirow{2}{*}{$1.4\!\!\times\!\!10^{7}$} & \multirow{2}{*}{$1.7\!\!\times\!\!10^{5}$} \\
      $\in(0, 42)$               &                                            &                                            &                                            &                                            &                                            \\
      Multiplicity               & \multirow{2}{*}{$5.1\!\!\times\!\!10^{5}$} & \multirow{2}{*}{$1.0\!\!\times\!\!10^{8}$} & \multirow{2}{*}{$2.7\!\!\times\!\!10^{6}$} & \multirow{2}{*}{$1.3\!\!\times\!\!10^{7}$} & \multirow{2}{*}{$1.5\!\!\times\!\!10^{5}$} \\
      $\in(40, 130)$             &                                            &                                            &                                            &                                            &                                            \\
      $E_{\rm leading\ neutral}$ & \multirow{2}{*}{$5.0\!\!\times\!\!10^{5}$} & \multirow{2}{*}{$9.2\!\!\times\!\!10^{7}$} & \multirow{2}{*}{$2.5\!\!\times\!\!10^{6}$} & \multirow{2}{*}{$1.2\!\!\times\!\!10^{7}$} & \multirow{2}{*}{$1.5\!\!\times\!\!10^{5}$} \\
      $\in(0, 41)$               &                                            &                                            &                                            &                                            &                                            \\
      $P_{T}$                    & \multirow{2}{*}{$4.3\!\!\times\!\!10^{5}$} & \multirow{2}{*}{$8.9\!\!\times\!\!10^{5}$} & \multirow{2}{*}{$1.4\!\!\times\!\!10^{6}$} & \multirow{2}{*}{$6.9\!\!\times\!\!10^{6}$} & \multirow{2}{*}{$1.3\!\!\times\!\!10^{5}$} \\
      $\in(20, 60)$              &                                            &                                            &                                            &                                            &                                            \\
      $P_{l}$                    & \multirow{2}{*}{$4.2\!\!\times\!\!10^{5}$} & \multirow{2}{*}{$1.9\!\!\times\!\!10^{5}$} & \multirow{2}{*}{$6.4\!\!\times\!\!10^{5}$} & \multirow{2}{*}{$3.0\!\!\times\!\!10^{6}$} & \multirow{2}{*}{$1.2\!\!\times\!\!10^{5}$} \\
      $\in(0, 50)$               &                                            &                                            &                                            &                                            &                                            \\
      $-\log_{10}(y_{23})$ & \multirow{2}{*}{$3.4\!\!\times\!\!10^{5}$} & \multirow{2}{*}{$1.5\!\!\times\!\!10^{5}$} & \multirow{2}{*}{$3.1\!\!\times\!\!10^{5}$} & \multirow{2}{*}{$1.1\!\!\times\!\!10^{6}$} & \multirow{2}{*}{$3.8\!\!\times\!\!10^{4}$} \\
      $\in(3.375, +\infty)$      &                                            &                                            &                                            &                                            &                                            \\
      $M_{\rm invariant}$        & \multirow{2}{*}{$2.6\!\!\times\!\!10^{5}$} & \multirow{2}{*}{$8.1\!\!\times\!\!10^{4}$} & \multirow{2}{*}{$6.2\!\!\times\!\!10^{4}$} & \multirow{2}{*}{$3.3\!\!\times\!\!10^{5}$} & \multirow{2}{*}{$2.5\!\!\times\!\!10^{4}$} \\
      $\in(110, 134)$            &                                            &                                            &                                            &                                            &                                            \\
      BDT                        & \multirow{2}{*}{$1.5\!\!\times\!\!10^{5}$} & \multirow{2}{*}{$1.2\!\!\times\!\!10^{4}$} & \multirow{2}{*}{$3.6\!\!\times\!\!10^{3}$} & \multirow{2}{*}{$6.3\!\!\times\!\!10^{3}$} & \multirow{2}{*}{$1.4\!\!\times\!\!10^{3}$} \\
      $\in(0.1, +\infty)$        &                                            &                                            &                                            &                                            &                                            \\

      \hline
    \end{tabular}
    
    }
\end{table}
 
The remaining events are then processed with toy MC to mimic the jet origin identification and the GBDT classifier, leading to the distribution shown in Fig.~\ref{fig:HssDistribution} in the letter.

\section{\label{appB}Appendix B: Comparative analyses of jet origin identification}

This appendix compares the performance of jet origin identification for different samples.
These samples are all full simulation samples using the CEPC baseline detector geometry and perfect lepton and charged hadron identification corresponding to the default scenario of particle identification.

\begin{figure}[!htbp]
    \centering
    \includegraphics[scale=0.5]{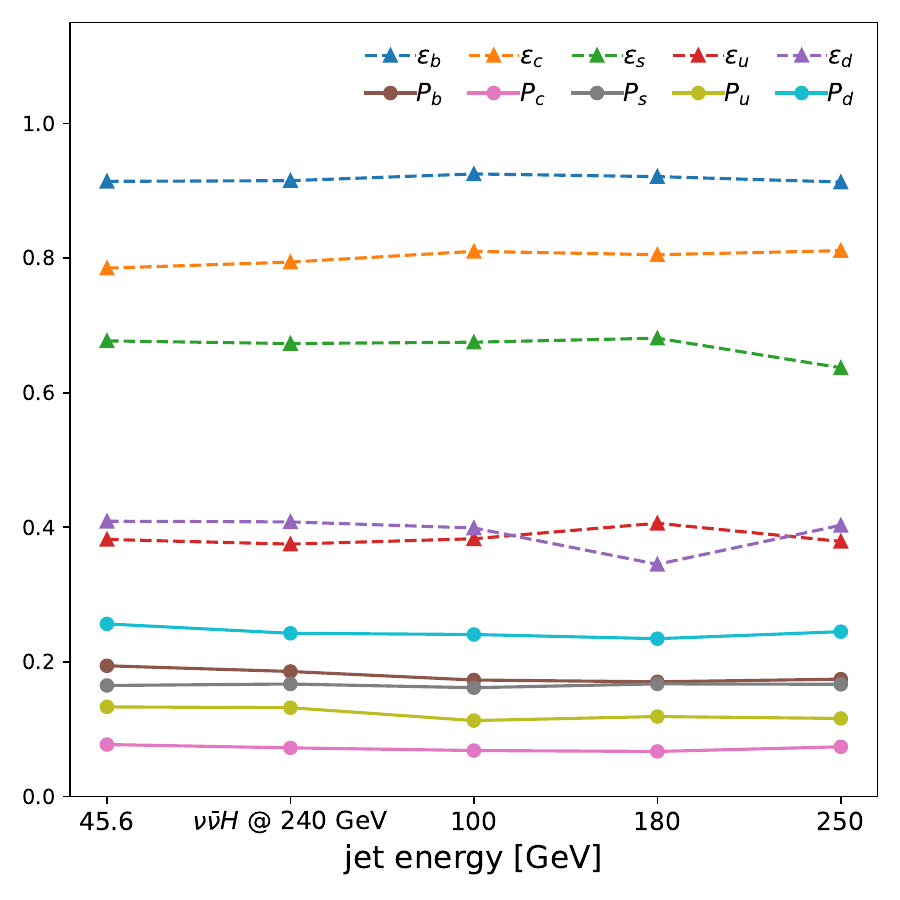}
    \caption{The jet origin identification performance: Jet flavor tagging efficiencies ($\varepsilon$) and charge flip rates ($P$) for various jet energies. 
    The error for each value is less than the per-thousand level.}
    \label{fig:energy}
\end{figure}

\begin{figure}[!htbp]
    \centering
    \includegraphics[scale=0.5]{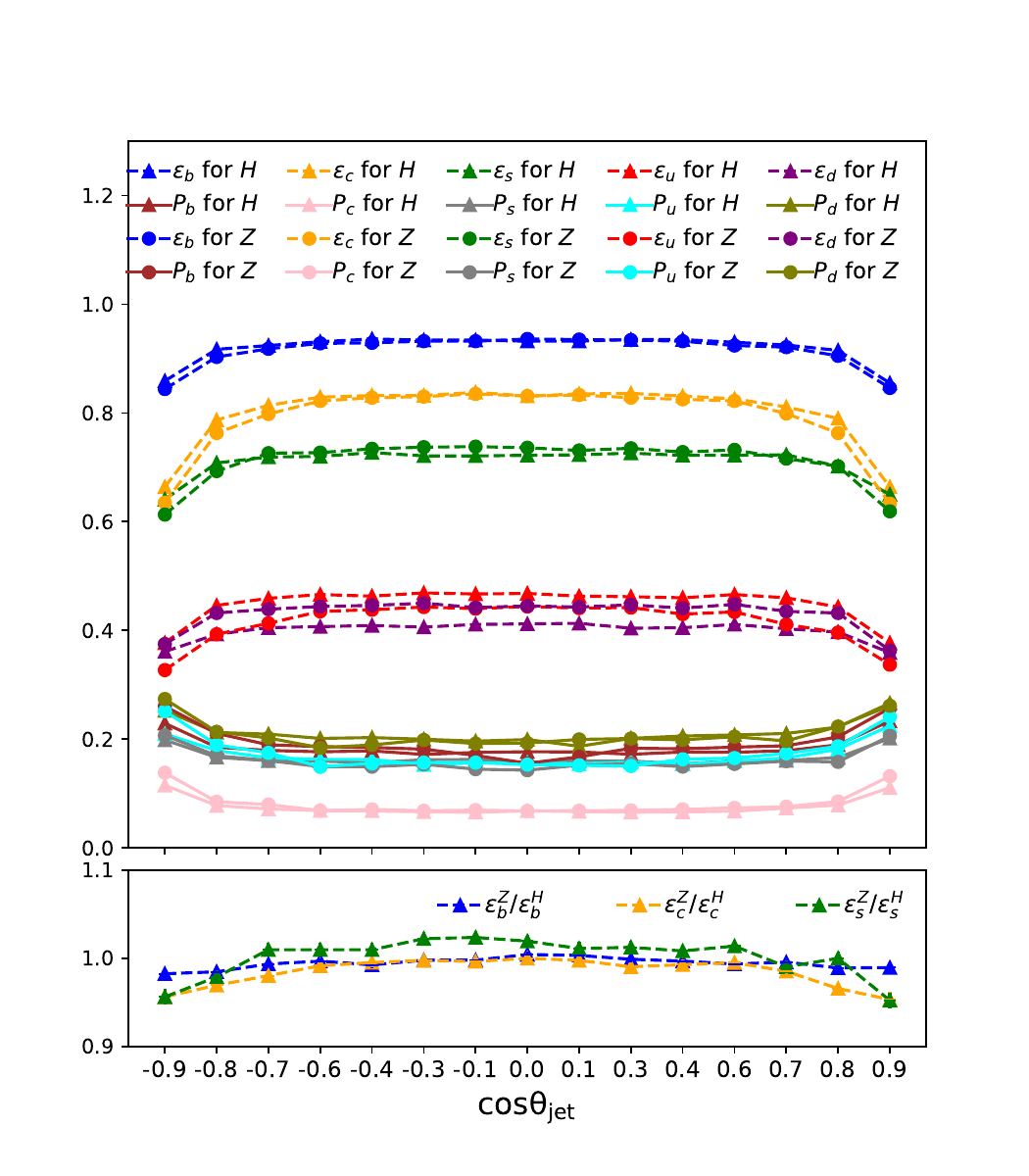}
    \caption{ 
    The jet flavor tagging efficiencies ($\varepsilon$) and charge flip rates ($P$) at different jet polar angles, 
    corresponding to both the $Z \rightarrow jj$ process at 91.2~GeV and the $\nu\bar{\nu} H, H \rightarrow jj$ process at 240~GeV. 
    The lower panel displays the ratios of flavor tagging efficiencies for $b$, $c$, and $s$ jets between these processes, showing the relative differences at the few-percent level, comparable to the statistical uncertainties.}  
    \label{fig:ERZH}
\end{figure}

\emph{1. Dependence on the jet energy and jet polar angle.} We extract the jet flavor tagging efficiencies and charge flip rates for various jet energies and polar angles.
On top of the $\nu\bar{\nu} H, H\to jj$ sample at 240~GeV center-of-mass energy, 
we simulate a Higgs boson at rest with changing mass, and the Higgs boson is forced to decay into a pair of jets.  
The Higgs boson mass is set to be 91.2, 200, 360, and 500~GeV, 
corresponding to jets with energies from 45.6 to 250~GeV.  
Fig.~\ref{fig:energy} shows the performance at different jet energies, where the extracted jet tagging efficiencies and charge flip rates are rather stable.
Fig.~\ref{fig:ERZH} shows the performance versus the jet polar angle, which is flat in the barrel region of the detector ($|\cos\theta| < 0.8$) and exhibits slight degradation in the endcap region.

\begin{figure}[!htbp]
    \centering
    \includegraphics[scale=0.5]{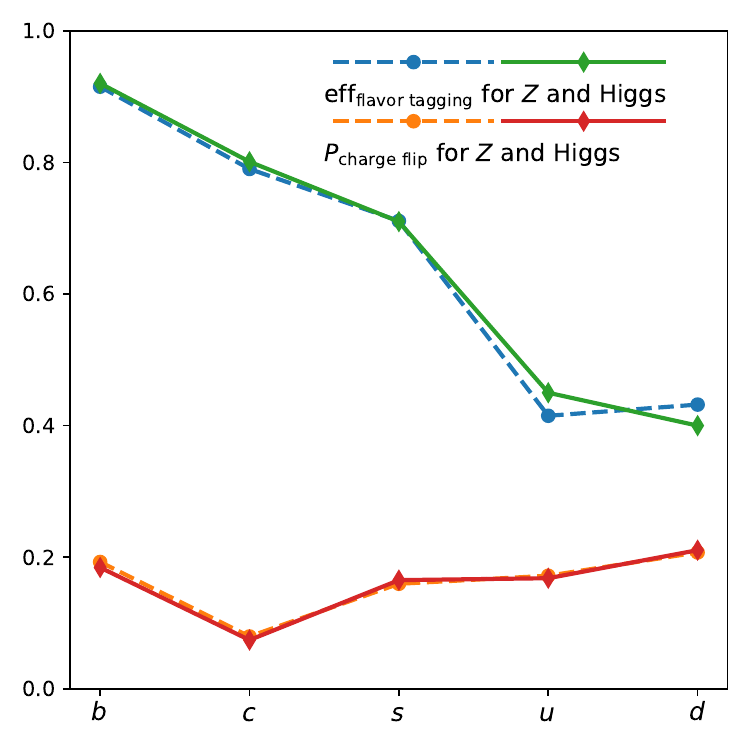}
    \caption{The comparison of flavor tagging efficiencies and charge flip rates between the $Z \rightarrow q\bar{q}$ process (dashed) at 91.2~GeV center-of-mass energy and the $\nu\bar{\nu} H, H \rightarrow q\bar{q}$ process (solid) at 240~GeV.
    This result is obtained using a 10-category classification for quarks, instead of an 11-category classification that includes a gluon category as presented in the main text.}
    \label{fig:ERHvsZ}
\end{figure}

\emph{2. Comparison between different physics processes.}
We compare the jet origin identification performance between the $Z \rightarrow q\bar{q}$ process at a center-of-mass energy of 91.2 GeV and the $\nu\bar{\nu} H, H \rightarrow q\bar{q}$ process at 240 GeV center-of-mass energy. 
We observe that the jet origin identification performance agrees  between these processes, especially in the fiducial barrel region of the detector for the flavor tagging performance of $b$, $c$, and $s$, see Fig.~\ref{fig:ERHvsZ}~and~\ref{fig:ERZH}.

It should be noted that, since the $Z$ boson does not decay into a pair of gluons, the gluon jet calibration is an open and interesting question, where dedicated QCD studies and usage of hadron collider data could be very helpful.

\emph {3. Comparison between different hadronization models. }
Jet origin identification uses directly the information of reconstructed final-state particles, while the hadronization process is responsible for generating final-state particles from initial quarks or gluons. 
The dependence of jet origin identification performance on the hadronization model is a natural concern.

\begin{figure}[!htbp]
    \centering
    \includegraphics[scale=0.5]{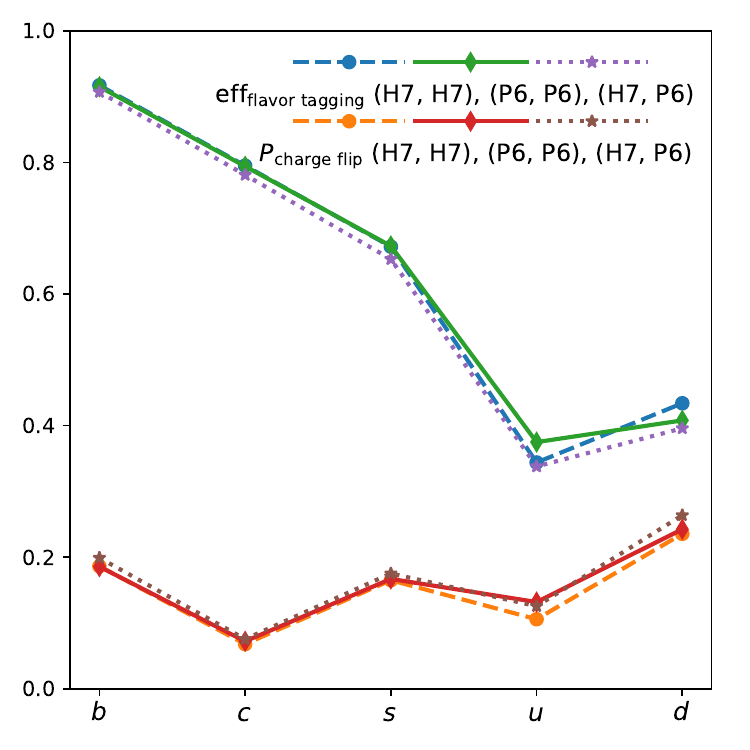}
    \caption{
    The performance comparison of flavor tagging efficiencies and charge flip rates of the $\nu\bar{\nu} H, H \rightarrow jj$ process at 240~GeV center-of-mass energy using Pythia-6.4 (P6) and Herwig-7.2.2 (H7).
    The legend brackets, i.e., (H7, P6) refers to the setup with training samples generated by Herwig and test samples generated by Pythia. 
    }
    \label{fig:P6H7}
\end{figure}

We compare the jet origin identification performance of samples derived from different hadronization models, namely Pythia-6.4 and Herwig-7.2.2~\cite{Bahr:2008pv, Bellm:2015jjp}. 
The predictions of the multiplicity of different final-state particles of these two hadronization models could be different by roughly 10\%~\cite{ruan2024Advanced}.
Figure~\ref{fig:P6H7} shows the performance with different training and test samples. To first order, the performance agrees between models, especially for $b$, $c$, and $s$ jets. 
The performance exhibits small but visible differences for $u$ and $d$ jets.

These comparative analyses show that the jet origin identification performance, especially for the heavy and strange quarks, is rather stable versus the jet kinematics (in the relevant energy range), different physics processes, and even different hadronization models.
The observed stability is vital for applying jet origin identification in real experiments.
Meanwhile, it is a critical and challenging task to determine and validate the fragmentation behavior of colored particles at a future Higgs factory.

\bibliographystyle{apsrev4-2} 
\bibliography{apssamp}

\end{document}